\documentclass[preprint,groupedaddress]{revtex4}%

\usepackage{graphicx}
\usepackage{amsmath}
\usepackage{epsfig}
\usepackage{color}
\usepackage{bm}
\usepackage[sort&compress]{natbib} 

\graphicspath{{Pics/}}

\begin{document}

\title{Exploring the nature of collisionless shocks under laboratory conditions 
}

\author{A. Stockem\(^1\)}
\email[Electronic address: anne.stockem@ist.utl.pt]{}
\author{F. Fiuza\(^{2}\)}
\author{A. Bret\(^{3,4}\)}
\author{R. A. Fonseca\(^{1,5}\)}
\author{L. O. Silva\(^1\)}
\email[Electronic address: luis.silva@ist.utl.pt]{}
\affiliation{\(^1\)GoLP/Instituto de Plasmas e Fus\~ao Nuclear - Laborat\'orio Associado, Instituto Superior T\'ecnico, Lisboa, Portugal\\
\(^2\)Lawrence Livermore National Laboratory, California\\
\(^3\)ETSI Industriales, Universidad de Castilla-La Mancha, 13071 Ciudad Real, Spain\\
\(^4\)Instituto de Investigaciones Energ\'eticas y Aplicaciones Industriales, Campus Universitario de Ciudad Real, 13071 Ciudad Real, Spain\\
\(^5\)ISCTE Instituto Universit\'ario Lisboa, Portugal}
 
\date{\today}

\begin{abstract}

Collisionless shocks are pervasive in astrophysics and they are critical to understand cosmic ray acceleration. Laboratory experiments with intense lasers are now opening the way to explore and characterise the underlying microphysics, which determine the acceleration process of collisionless shocks. We determine the shock character -- electrostatic or electromagnetic -- based on the stability of electrostatic shocks to transverse electromagnetic fluctuations as a function of the electron temperature and flow velocity of the plasma components, and we compare the analytical model with particle-in-cell simulations. By making the connection with the laser parameters driving the plasma flows, we demonstrate that shocks with different and distinct underlying microphysics can be explored in the laboratory with state-of-the-art laser systems.
\end{abstract}



\maketitle


One of the most important problems in astrophysics is the acceleration of charged particles to very high energies, e.\,g.\ to explain the cosmic ray spectrum \cite{SS76,CB11}, where shock acceleration is a very promising model for the power-law dependence at high energies, or the jet emission of particles in gamma-ray bursts \cite{P05}. The acceleration process in collisionless shocks is determined by the underlying microphysics and depends on the shock character, i.\,e.\ the electromagnetic fields are responsible for sustaining the shock transition, since in collisionless shocks the dissipation mechanism is determined only by the fields in the shock front and not by the inter-particle collisions since the mean free path is much larger than the shock front. For instance, cold electron distributions can lead to electromagnetic shocks \cite{S08,MF09}, while low flow velocities lead to electrostatic shocks \cite{FS70}, but the transition between the two different regimes and the parameter space where the different field structures dominate have not been defined yet.

Electromagnetic shocks are mediated by Weibel-type instabilities \cite{W59,F59} with the shock front transition determined by magnetic fields on the ion scale \( (\simeq c/\omega_{pi}) \). They usually appear in astrophysical scenarios, e.\,g.\ in the outflows of gamma-ray bursts, active galactic nuclei or pulsar wind nebulae. Since instabilities of the Weibel-type are very efficient in highly anisotropic plasmas, low thermal and high fluid velocities enhance the shock formation process \cite{SF02}. The study of electromagnetic shocks in the laboratory is difficult because of the long shock formation time at low fluid velocities \cite{BS13}, which are now obtained experimentally \cite{PR12,KR12}, or the need to drive relativistic speeds, which requires very intense lasers \cite{FF12}.

Electrostatic shocks, on the other hand, are a consequence of non-linear wave steepening and wave breaking of ion-acoustic modes and they are more relevant for laboratory experiments, e.\,g.\ in laser-plasma interactions \cite{HT12,AD13}, fast ignition \cite{VS12} and lately also for medical applications \cite{LA07} as alternative to standard laser-acceleration schemes \cite{MF08,R08,MB13}. The shock formation process is enhanced in plasmas with high mass and temperature ratios between the negatively and positively charged plasma components \cite{FS70,SM06}. Theoretical estimates are usually one-dimensional and neglect the importance of electromagnetic modes which are clearly associated with the multi-dimensional features of the problem.

Shocks naturally appear from the collision of two plasma slabs and are of electromagnetic character if transverse electromagnetic modes drive the shock formation process or of electrostatic character if longitudinal instabilities dominate. However, the shock character can change on long time scales due to competing wave processes, which has been phenomenologically identified in numerical simulations \cite{SF13spie,TC13}. The shocks will have different microphysics and lead to distinct spectra of accelerated particles. It is thus necessary to understand the dependence of the character of the shock on the flow to design laboratory experiments to clarify the underlying physics and to connect with astrophysics observations.

In this paper, we provide a theoretical prediction of the dominant shock character for the two main parameters characterising non-magnetised plasma flows. We determine regimes depending on the upstream parameters, and identify a region, where the transition from initially electrostatic shocks to electromagnetic shocks occurs. The theoretical findings are confirmed with particle-in-cell simulations and the transition between the different regimes is illustrated. Connections with possible experiments are made and we show that these distinct regimes can already be explored in the laboratory.


\section*{Results}
Our starting point is the analytical description of electrostatic shocks. This is a classical problem \cite{FS70,FF71,SM06,S66,S72,TK71,MJ69,S86} for which there are analytical solutions, in 1D, for the shock solution. We will use this solution to verify under which conditions the growth rate of electromagnetic Weibel modes associated with the distribution function is larger than the inverse electrostatic shock formation time. We first observe that an electrostatic shock solution exists.
Applying Poisson's equation to connect the electrostatic potential and densities of the upstream and downstream shock regions and treating the potential as a harmonic oscillator in the so-called Sagdeev potential \cite{S66} provides the valid shock solutions, which are limited by a maximum Mach number. The criterion still holds in the relativistic generalisation, and thus an electrostatic shock is formed if the relativistic Mach number \(M :=u_0/u_s \lesssim 3.1 \) with \(u_0 = \beta_0 \gamma_0\) the proper upstream velocity, \(u_s = \beta_s \gamma_s\) and \(\beta_s = \sqrt{k_B T_e / m_i c^2}\) the ion sound speed with \(T_e\) the initial electron temperature \cite{FS70,SF13}. 
This can be expressed as \(k_B T_e / m_e c^2 = ((3.1/u_0)^2+1)^{-1} m_i /m_e\) and establishes a first domain of the \(u_0 - T_e\) parameter space where shocks can be explored. From the growth rate for longitudinal modes for cold ions \cite{KT10}, the shock formation time is estimated as \(t_{sf} \approx 5 \omega_{ES}^{-1} = 10 \gamma_0^{3/2} \omega_{pi}^{-1} \), which we have confirmed in a series of electrostatic shock simulations. 

We now analyse the stability properties of the electrostatic shock solutions regarding the growth of transverse electromagnetic modes. We assume that the particle distribution after electrostatic shock formation is given by the model described in \cite{S72}, with the corresponding generalisation for relativistic beams \cite{SF13}. All quantities are given in the laboratory frame, which corresponds to the rest frame of the downstream population. The upstream populations of free streaming electrons and ions are affected by the shock potential, leading to a population of free streaming and trapped electrons downstream. Throughout the paper the ion populations are kept cold and we assume that the impact on their distribution is negligible at this stage. We consider here the simplest case of equal density and temperature ratios of the upstream and downstream populations, so that the free electron distribution is given by a drifting Maxwell-J\"uttner distribution \(f_{e,r\pm}(\gamma, u_x) = n_{0,r} \left(\mu /2 \pi \right)^{3/2}  \, \exp \left\{-\mu \left[ \gamma_0 \left( \gamma - \varphi \right) -1 \pm u_0 \sqrt{(\sqrt{1+u_x^2}-\varphi)^2-1} \right] \right\}\), where \(u_0 = \sqrt{1+ \gamma_0^2} \) is the proper bulk velocity of the upstream in the shock frame, with propagation direction along the \(x\) axis, \(\mu =  m_e c^2 / k_B T_e\) is the thermal parameter and \(\varphi= e \phi / m_e c^2\) the electrostatic potential. This distribution is valid if the kinetic energy exceeds the potential energy, leading to the condition \( \gamma > \gamma_c =  1 + \varphi\) for the Lorentz factor, where \(f_{e,r+}\) is valid for \(u_x < -u_c\) and  \(f_{e,r-}\) for \(u_x > u_c\) with \(u_c = \sqrt{\gamma_c^2-1}\). In the non-relativistic limit, this distribution is given by \(f_{e,\pm} = n_0  \left(\mu /2 \pi \right)^{3/2}  \exp\{-\mu[( \sqrt{\beta_x^2- 2 \varphi} \pm \beta_0)^2+ \beta_y^2+\beta_z^2]/2\} \) with  \( n_0 = \left[e^{\mu \varphi} \textrm{erfc} \sqrt{\mu \varphi} + 2  \sqrt{\mu \varphi/\pi} \exp\left[-\mu \beta_0^2 / 2\right]  \right]^{-1}\). The electrons are trapped if their kinetic energy is lower than the energy of the electrostatic potential, \(\gamma < \gamma_c\). In analogy to the non-relativistic approach, where the trapped population is described with a flat-top profile \cite{S72,SM06}, we formulate the relativistic distribution of trapped particles as  \(f_{e,rt} = n_{0,r} \left(\mu /2 \pi \right)^{3/2}    \exp \left\{-\mu \left[ \gamma_0 \gamma_\perp -1 \right] \right\}\) with \(\gamma_\perp = \sqrt{1+u_y^2+u_z^2}\), which goes to \(f_{e,t} \approx n_{0}  \left(\mu /2 \pi \right)^{3/2}  \exp\{-\mu (\beta_0^2+\beta_y^2+\beta_z^2)/2\} \) in the non-relativistic approximation.

The dispersion relation of electromagnetic filamentation/Weibel modes is calculated, which is not shown here because of its length, and a numerical solution can be found for the general (relativistic) case. In the non-relativistic approximation, \(v_0 \ll c\), the dispersion relation can be approximated by
\begin{equation}\label{eq1}
	k^2c^2-\omega^2+ \omega_{pe}^2 \left[ 1- V(\varphi ) \left[  1+ \frac{\omega}{kc} \sqrt{\frac{\mu}{2}} \textrm Z \left(  \frac{\omega}{kc} \sqrt{\frac{\mu}{2}}  \right) \right] \right] =0,
\end{equation}
with frequency \(\omega\), wave number \(k\), plasma dispersion function \(Z\) \cite{FC61} and \(V(\varphi )= n_0  \left \{ e^{\mu \varphi} \textrm{erfc} \sqrt{\mu \varphi} + 2 \sqrt{\mu \varphi / \pi} + \frac{4}{3} \sqrt{\mu^3 \varphi^3 / \pi } \exp\left[-\mu \beta_0^2 / 2\right]  \right\}\), resembling the well-known dispersion relation for the Weibel instability \cite{W59}, where the potential-dependent term plays the role of the anisotropy parameter \(A\), with \(V(\varphi) = 1 + A = v_{th\parallel}^2/v_{th\perp}^2\). This anisotropy is introduced by the distortion of the distribution function in the longitudinal direction due to the electrostatic field in the shock. The growth rate is defined as the imaginary part of the frequency \(\sigma = \Im(\omega)\). In the case \(\mu \varphi \gg 1 \) and \(\beta_0 \ll 1\), the approximation \(V (\varphi) \approx 1+\mu (2\varphi/3 - \beta_0^2/2)\) can be found. For small arguments of the \(Z\) function, which is the case for very low fluid velocities and thermal velocities the growth rate is given by \( \sigma \approx k c \sqrt{2/\mu \pi} \left[ 1-(k^2 c^2 + \omega_{pe}^2)/(\omega_{pe}^2V(\varphi )) \right]
\) with \(k_0^2c^2 = \omega_{pe}^2(V(\varphi)-1)/3\) and a maximum growth rate 
\begin{equation}\label{nonrel}
	\sigma_{max} \approx \sqrt{\frac{1}{\pi \mu}}\frac{\omega_{pe}}{V(\varphi)} \left( \frac{2}{3} (V(\varphi )-1)  \right)^{3/2}.
\end{equation}
The Weibel modes are considered to be relevant if their time scale \( t_W \simeq \sigma^{-1} \) is comparable with the electrostatic shock formation time scale \(t_{sf} = 10 \,  \gamma_0^{3/2} \omega_{pi}^{-1}\). The growth rate depends on the potential across the shock front, and decreases from the upstream to the downstream. In order to define parameter regimes specifically, it is assumed that ion reflection from the potential has just set in, which is equivalent to consider \( \varphi_{max} = m_i / m_e (\gamma_0 -1 )\). The details about the dependence of the growth rate of the Weibel modes across the shock front and with the potential will be presented elsewhere.

\begin{figure}[h!]
\includegraphics[width=9cm]{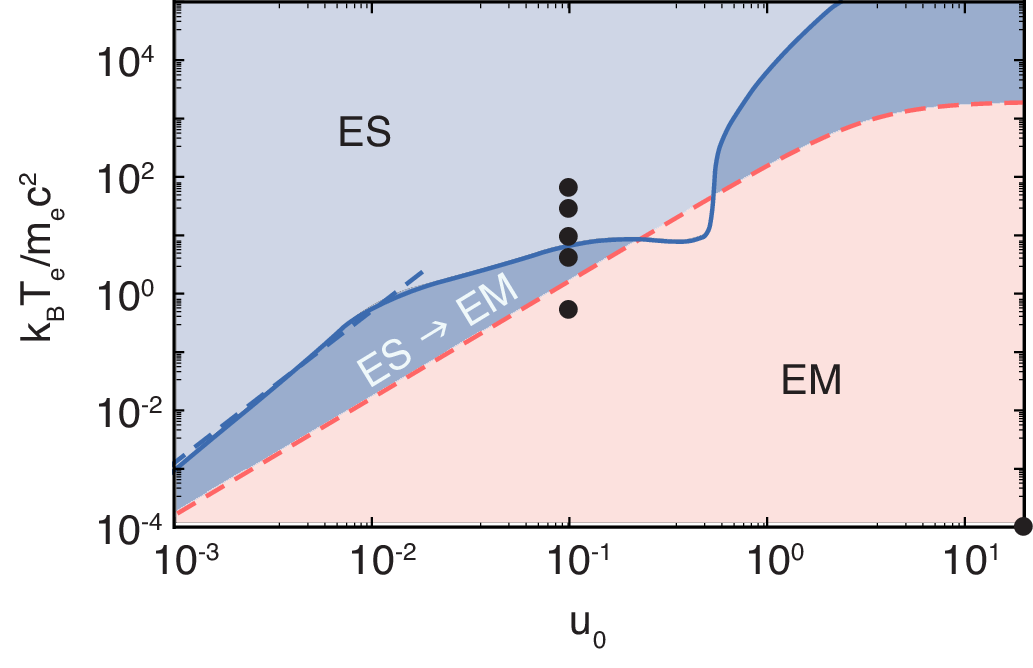}
\caption{Definition of electrostatic, electromagnetic and transition regimes depending on input parameters \( k_B T_e / m_e c^2\) and proper flow velocity \(u_0 = \beta_0 \gamma_0\). The electrostatic shock formation condition \cite{FS70} (red dashed line) limits the parameter space of electromagnetic shocks (EM). The blue line represents the condition \(t_{W} = t_{sf}\) with the nonrelativistic approximation in dashed, separating the region of purely electrostatic shocks (ES) and the transition region (ES \(\rightarrow\) EM). The black dots represent the sub-set of simulation parameters discussed in the paper.}\label{fig1}
\end{figure}

As a result of the theoretical analysis, parameter regimes of the dominating electrostatic or electromagnetic shock character can be defined, which are summarised in Figure \ref{fig1}. The region of purely electromagnetic shocks (EM), which is determined by the electrostatic shock formation condition \(u_0 \lesssim 3.1 \sqrt{k_B T_e / m_i c^2}\) \cite{FS70}, and purely electrostatic shocks (ES) are separated by a transition region (ES \(\rightarrow\) EM), where the growth rate of the Weibel instability is larger than the shock formation time scale. 

The transition between the different regimes, shown in Figure \ref{fig1}, can be observed in particle-in-cell simulations with details given in the Methods section and we discuss in detail the three representative cases \(k_B T_e / m_e c^2 = 10\), 3.5, 0.5 with proper flow velocity \(u_0 = 0.1\). The role of the cold ion-ion instability has been addressed in \cite{FF12}.

In the electrostatic regime (ES), and for \(k_B T_e / m_e c^2 = 10\), after 12 \(\omega_{pi}^{-1}\) the potential energy in the shock front exceeds the kinetic energy of the upstream protons and goes into a steady state with a downstream to upstream density ratio \(n_d/n_0 = 2.7\) and shock speed \(v_{sh} = 0.048 \,c\) and a second peak behind the shock propagating with \(0.041 \,c\), in agreement with \cite{SF13}. The potential energy is much larger than the proton kinetic energy, so that most of the upstream protons are reflected back into the upstream and a quasi monoenergetic spectrum is created, clearly illustrating the distinct features of ion acceleration in electrostatic shocks.

The electromagnetic shock (EM) evolves on much longer time scales and protons thermalise only on thousands \(\omega_{pi}^{-1}\). The two counter propagating plasma slabs just interpenetrate and overlap each other, until the filamentation instability starts to slow down the protons. The electrostatic potential across the shock is not strong enough to reflect them and ion acceleration occurs on a much longer time scale on a Fermi-like process \cite{S08,MF09}.

\begin{figure}[h!]
\includegraphics[width=11cm]{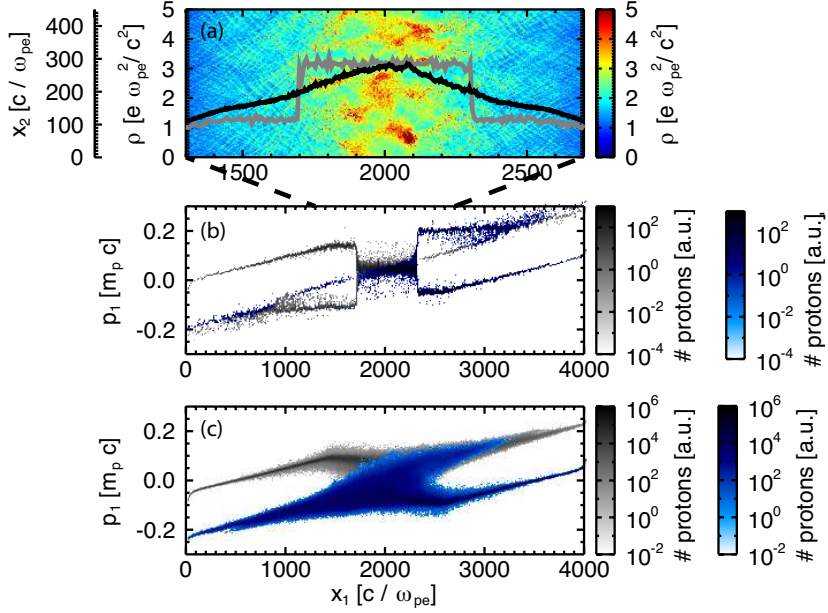}
\caption{(a) Density in the ES/EM transition case \(k_B T_e / m_e c^2 = 3.5\) and comparison of 1D (grey) and 2D (black) simulation. Proton phase spaces in 1D (b) and 2D (c). All are shown for \(t\omega_{pe} = 10^4\).}\label{fig23}
\end{figure}

In the transition region (ES \(\rightarrow\) EM), we observe two stages. In the first stage, a strong electrostatic field appears and a shock is formed, showing the characteristics of an electrostatic shock. The potential energy is on the order of the upstream kinetic energy and most of the protons are reflected, such that a quasi monoenergetic spectrum of ions is formed. At approximately 12 \(\omega_{pi}^{-1}\) a second potential appears at the centre of the box and propagates outwards with a decreased potential jump of 0.6 \(E_{kin,p}\). Only a fraction of the upstream protons is reflected at the shock front and the proton spectrum becomes more diffuse, see Figure \ref{fig23}. The protons start to thermalise in the downstream region. The shock is now in a quasi-steady state, after \(\approx 50 \, \omega_{pi}^{-1}\), with a decreased shock velocity \(0.022 \,c\) and the density ratio increases to approximately 3 \cite{SF12}. The potential jump becomes less sharp and the shock front region extends to larger spatial scales with the same order of potential difference between the upstream and downstream region. The shock front is now determined by a mix of transverse and longitudinal modes, but without the mono energetic features of the ion spectrum.

\begin{figure}[h!]
\includegraphics[width=12cm]{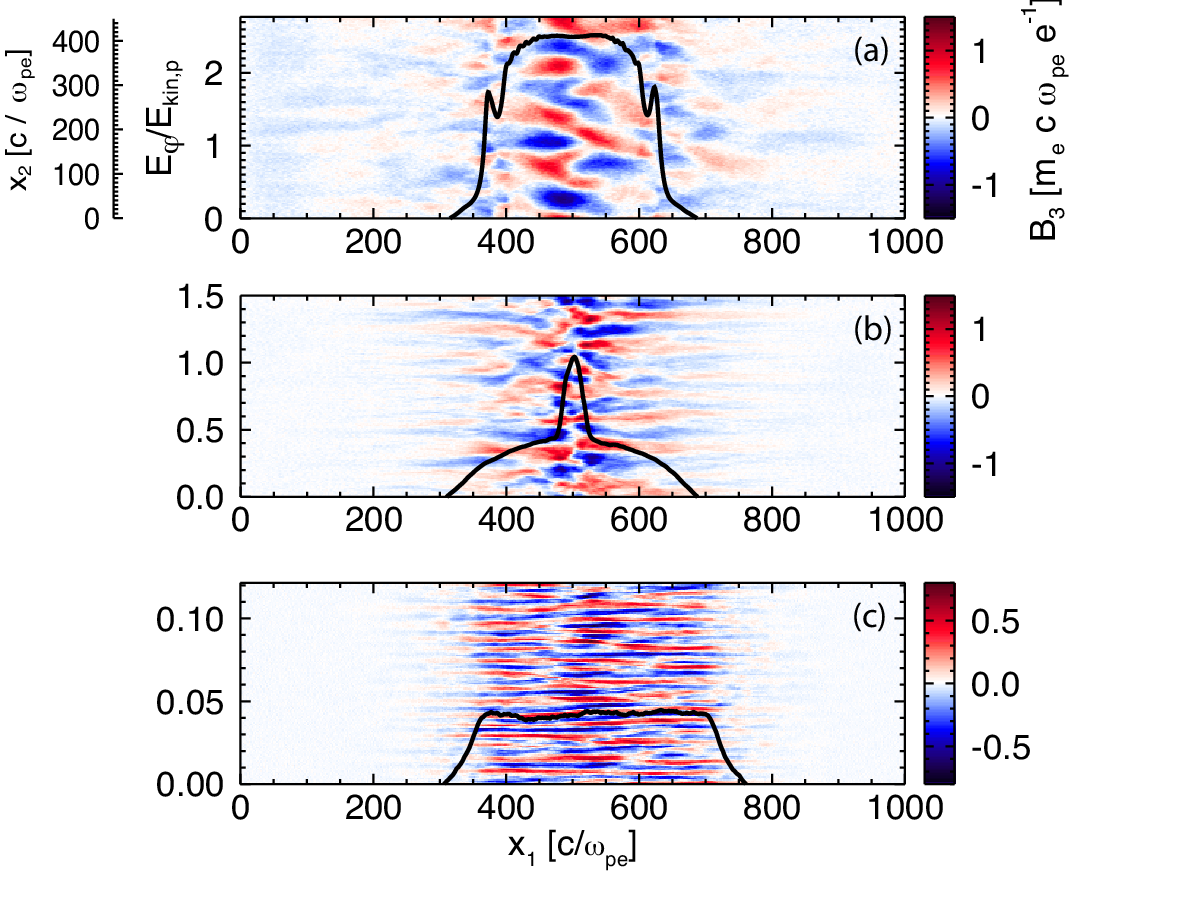}
\caption{Magnetic field \(B_3\) in the entire simulation box at early stage of shock formation, \(t \omega_{pe} = 2500\), for \(k_B T_e / m_e c^2  = 10\) (a), 3.5 (b) and 0.5 (c). Overplotted is the electrostatic potential normalised to the ion kinetic energy \(E_\varphi / E_{kin,p}\) showing the transition from the electrostatic (a) to the electromagnetic case (c).}\label{fig2}
\end{figure}

The transition of the shock formation process between the different regimes is illustrated in Figure \ref{fig2} at \(t\omega_{pe} = 2500 \). We note that a perpendicular magnetic field \(B_3\) appears in all three regimes, but the character changes significantly. In the case of an electrostatic shock (fig. \ref{fig2} a) the spatial scale is large with a wavelength \(\lambda = c / \omega_{pi}\) in the linear phase of magnetic field generation. In the transition case (b) the wavelength is decreased to \(0.6 \, c / \omega_{pi}\) and a more ordered structure appears with filaments aligned with the shock propagation direction \(x_1\), which resembles more the ordered structure of the electromagnetic case (c), where the characteristic wavelength is \(0.2 \, c / \omega_{pi}\), closer to the length scale for electron filamentation. At this stage, the shock has not fully formed yet, but the change in the electron dynamics, which are determining the shock formation process, is clearly visible.

Also the shape and strength of the electrostatic potential are closely connected with the shock character. In the electrostatic case, a potential jump is generated as expected from electrostatic theory \cite{S72}, showing oscillations that indicate particle trapping in the downstream. In the steady state, the energy of the electrostatic potential is \(E_\varphi = 1.6 \, E_{kin,p}\), i.\,e.\ high enough to reflect the upstream ions.

\begin{figure}[h!]
\includegraphics[width=10cm]{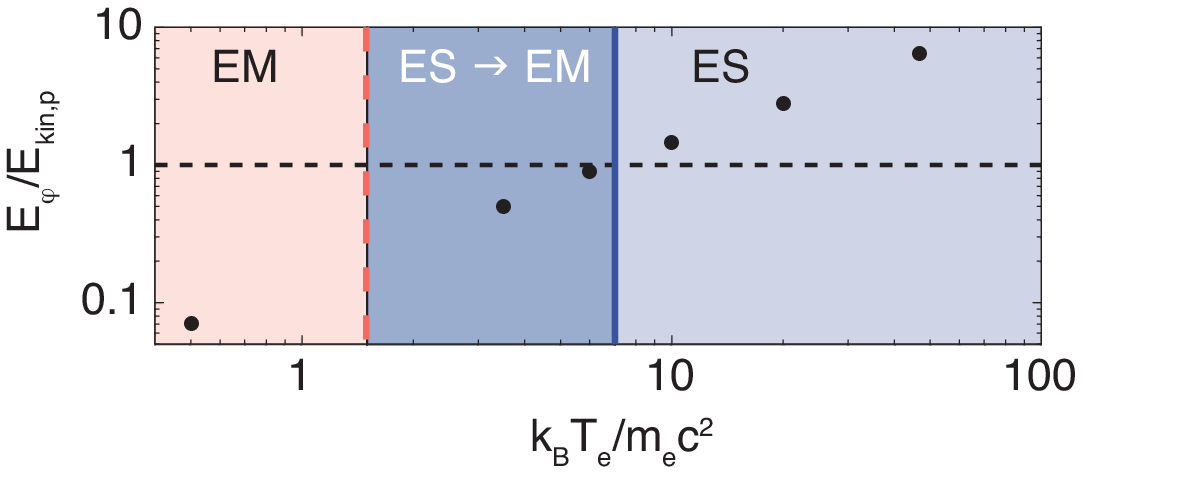}
\vspace{-12pt}
\caption{Potential energy over ion kinetic energy against \(k_B T_e /m_e c^2  \) measured from simulation data. The horizontal dashed line represents the ion reflection condition \(E_{\varphi} > E_{kin,p}\). The regions EM, ES \(\rightarrow\) EM, ES are taken from Figure \ref{fig1} for \(u_0 = 0.1\).}\label{fig3}
\end{figure}

In the transition case, a first potential develops which smoothens out while a second potential appears from the center of the box (see fig.\ \ref{fig2}b). It shows no signs of particle trapping, i.\,e.\ oscillations in the downstream potential. The differences in mass and temperature between protons and electrons also lead to a potential in the electromagnetic case \(E_{\varphi} = 0.07 E_{kin,p}\), which is so small that almost no protons are reflected from it. The potential jumps for the different scenarios are shown in Figure \ref{fig3}. The ability of ion reflection is increased with the electron temperature. We observe a matching with the regimes defined in Figure \ref{fig1} since the ion reflection condition \(E_\varphi / E_{kin,p} > 1\) coincides with the threshold for purely electrostatic shocks (ES) at around \(k_B T_e / m_e c^2 \approx 7\) for \(u_0 = 0.1\).

\begin{figure}[h!]
\includegraphics[width=15cm]{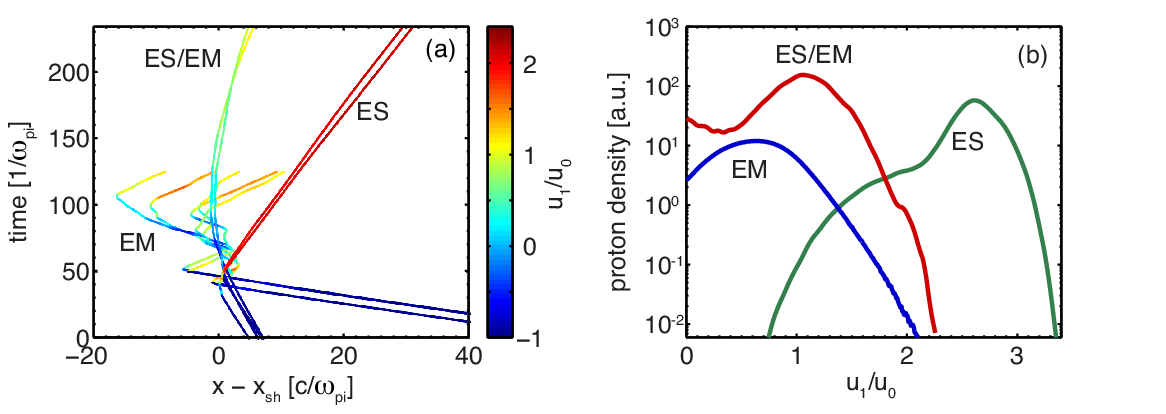}
\caption{(a) Proton trajectories for the three different regimes defined in Fig.\ \ref{fig1} relative to the position of the shock front \(x_{sh}\) and colour coded with the longitudinal momentum \(u_1 = p_1 / m_p c\), which was normalised to the momentum of the upstream fluid \(u_0\), and (b) proton spectra in the downstream and shock front region after the steady state of shock formation.}\label{fig4}
\end{figure}

An analysis of the particle dynamics in the field structure of the shocks supports our findings. To capture the motion in a fully developed shock, here, the electromagnetic case (EM) is presented by a run with a reduced mass ratio and relativistic upstream fluid velocity. Figure \ref{fig4}(a) shows typical proton trajectories for the three different regimes. In the case of an electrostatic shock, the upstream protons are picked up by the shock front and reflected back into the upstream with velocity \(v_{i,refl} \approx v_0 + 2 v_{sh}\) in the non-relativistic case. These particles gain the highest energies for the time scales in our comparison. In electromagnetic shocks, the particles gain energy by scattering off the electromagnetic turbulences in the shock front region. This scattering process can be well observed in fig.\ \ref{fig4}(a). In the transition region (ES/EM), the potential is not strong enough to reflect the fast upstream protons, and they are decelerated while approaching the shock front. Only the slowest particles are scattered back into the upstream region with velocities much smaller than \(v_{i,refl}\). The accelerated proton spectra in Figure \ref{fig4}(b) thus show fundamental differences. While in the electrostatic case, the velocities are highest and centred with a low spread around the value \(v_{i,refl}\), in the other two regimes the accelerated spectra are broad and centred around lower values.

\section*{Discussion}
In order to make the connection with laser experiments, we now connect the parameters determining the shock character with the laser parameters in laser-solid interactions. We assume that two symmetric counter propagating flows are created by irradiating a plasma target with a laser. This can be done by either using two laser-irradiated foils in order to create two counter streaming flows \cite{TK08}, or by using a single high-intensity laser that accelerates the plasma into the target and the return current provides the required counter-streaming conditions for the development of the instability \cite{FF12}. In this last scenario, the temperature of the plasma flow \(T_e\) can be connected with the laser via the ponderomotive scaling of \(T_e\) with the laser normalised intensity \(a_0\) via \(k_B T_e / m_e c^2 = \sqrt{1+a_0^2}-1\) \cite{WK92}. For the flow velocity, one can use the hole boring velocity as obtained in \cite{WK92}, \(\beta_0 = \sqrt{n_c m_e / (2 n_0 m_p)} a_0\) with critical density \(n_c\). We observe that the physics of ponderomotive heating and hole boring is quite complex but the approximate scaling laws can provide a guide to the required laser parameters. During the characteristic time of the laser pulse \(\tau \), which should be on the order of a few times the electrostatic shock formation time \(t_{sh} = 5 \lambda_0/(\pi c)  \sqrt{n_c m_p / n_0 m_e} / (1-a_0^2 n_c m_e/2 n_0 m_p)^3\) with \(\lambda_0\) the laser wavelength, the plasma slabs overlap for \(L = 2 \, v_0 \, \tau\), which defines the minimum length for the target thickness. For geometrical reasons and in order to guarantee the shock front to be as plane as possible, the laser focal spot size should be large compared with the shock front thickness, \(w_0 \gg c/\omega_{pe} = \lambda_0 / (2\pi) \sqrt{n_c / n_0}\).
Simulations show that the hole boring process is connected with a non-regular pile up of the density \cite{LY08}. For simplicity, we assume a homogeneous initial density distribution with \(n_0 / n_c = 1-50 \) and plan to address the modifications due to a realistic smoothly growing density profile in a future work. Assuming a laser wavelength \(\lambda_0 = 1\, \mu\)m and \(a_0 \approx 5 -50\), which is in the range of state-of-the-art laser systems and ongoing laser experiments, the shock formation time is on the ps time scale with a target size of \(10-100\, \mu\)m. We can expect \(T_e\) in the range 2-25 MeV and \(\beta_0 \gtrsim 0.1\), thus capable of exploring the transition. 

Starting from the seminal analytical solution of an electrostatic shock, we have calculated if the system is stable to Weibel or filamentation modes. Due to the distorted electron distribution of free and trapped particles in the downstream of the shock, an anisotropy arises that enhances the excitation of electromagnetic modes. Depending on the initial electron temperature and fluid velocity, parameter regimes were identified in which the shock solution will be purely electrostatic, electromagnetic or shows a transition between the two regimes.

The transition between the different shock characters has been confirmed by particle-in-cell simulations. In the transition regime, a change of the shock character appeared with a shock formation time characteristic for electrostatic shocks. The appearance of a magnetic field was observed in all three cases, with a significantly different character and different temporal and spatial scales. For electrostatic shocks, the temporal and spatial scales of the magnetic field are \(10\,  \omega_{pi}^{-1}\), \(c/\omega_{pi}\); in the transition region it grows on \(5\,  \omega_{pi}^{-1}\), \(0.6 \, c/\omega_{pi}\), while for electromagnetic shocks \(\omega_{pi}^{-1} \), \(0.2 \,c/\omega_{pi}\). Furthermore, we found the reflection condition as an alternative criterion for distinguishing between the different regimes, since it matches with the boundary of the pure electrostatic shock regime. The dominant acceleration process in the different regimes, a function of the fields mediating the shock, determines the acceleration spectra of the protons. While in the electromagnetic case a broad spectrum is generated, highest particle velocities with rather small velocity spread are obtained in the electrostatic case. By using standard scaling for the electron temperature and hole boring velocity with the laser intensity, we showed that the relevant regimes discussed in this paper can already be studied with existing laser systems.

\section*{Methods}
Two-dimensional numerical simulations were carried out with the particle-in-cell code OSIRIS (ref.\ \cite{F02,F08}) modelling the shock formation due to two counter streaming plasma flows of electrons and ions. Two sets of simulations have been performed with the fluid velocity of the plasma slabs \(v_0= \pm 0.1 \,c\) and thermal parameters \(k_B T_e / m_e c^2 = 0.5\), 3.5, 10, 20 and 50 of a relativistic Maxwell-Juettner distribution. High resolution runs up to \( t_{max} = 2500 \, \omega_{pe}^{-1}= 58 \, \omega_{pi}^{-1}\) were performed with a simulation box covering \(L_x = 1000\, c/\omega_{pe}\) and \( L_y = 450\, c/\omega_{pe}\) and a spatial resolution \(\Delta x = \Delta y = 0.1 \, c / \omega_{pe}\) to resolve the Debye length. All quantities are normalised to \(\omega_{pe} = \sqrt{4 \pi n_{0,e} e^2/m_e}\), the non-relativistic electron plasma frequency with respect to the upstream electron density \(n_{0,e}\). In the second set of simulations, the shock evolution was followed up to \(t_{max} = 10^4 \, \omega_{pe}^{-1}= 233 \, \omega_{pi}^{-1}\) with a decreased resolution \(\Delta x = \Delta y = 0.5 \, c / \omega_{pe}\) and the simulation box covering \(L_x = 8000\, c/\omega_{pe}\) and \( L_y = 200\, c/\omega_{pe}\). All simulations were performed with a realistic proton to electron mass ratio \(m_p/ m_e = 1836\). A cubic interpolation scheme and 9 particles per cell and per species were used in all simulations. To obtain the typical particle trajectories in the electromagnetic regime, a simulation with reduced mass ratio \(m_i / m_e = 50\), relativistic Lorentz factor \(\gamma_0 = 20\) and cold distributions \(k_B T_e / m_e c^2 = 10^{-4}\) was performed. The simulation box dimensions were \(L_x = 4000\, c/\omega_{pe}\) and \( L_y = 320\, c/\omega_{pe}\) with resolution \(\Delta x = 0.1 \sqrt{\gamma_0} \, c/\omega_{pe}\),  \(\Delta t = 0.05 \sqrt{\gamma_0} /\omega_{pe}\) and \(t_{max} = 4000 / \omega_{pe} = 126 \, \sqrt{\gamma_0}/ \omega_{pi}\).


\section*{Acknowledgements}
This work was supported by the European Research Council (ERC-2010-AdG Grant 267841), FCT (Portugal) grants SFRH/BPD/65008/2009, SFRH/BD/38952/2007, and PTDC/FIS/111720/2009 and projects ENE2009-09276 of the Spanish Ministerio de Educaci\'on y Ciencia. We would like to acknowledge the assistance of high performance computing resources provided by PRACE on Jugene and SuperMuc at the Leibniz-Rechenzentrum based in Germany. Simulations were performed at the IST cluster (Lisbon, Portugal), the Jugene/Juqueen and SuperMuc supercomputers (Germany).
We thank Laurent Gremillet and Charles Ruyer for fruitful discussions on this work.

\section*{Author Contributions}
AS performed the calculations and numerical simulations in collaboration with LOS and RAF on the technical part of the numerical simulations. All authors (AS, FF, AB, RAF, LOS) contributed to the manuscript preparation. LOS provided overall guidance to the projects.

\section*{Additional information}
The authors declare no competing financial interests.

\section*{Dispersion relation of electromagnetic modes}

The relativistic distribution function
\begin{equation}
	f_{re} = n_{r0} %
	\begin{cases}
	\exp \left\{-\mu \left[ \gamma_0 \left( \gamma - \varphi \right) -1 + u_0 \sqrt{(\sqrt{1+u_x^2}-\varphi)^2-1} \right] \right\}	 & \quad u_x < - u_c\\
	\exp \left\{-\mu \left[ \gamma_0 \gamma_\perp -1 \right] \right\} & \quad |u_x| \leq u_c\\
	\exp \left\{-\mu \left[ \gamma_0 \left( \gamma - \varphi \right) -1 - u_0 \sqrt{(\sqrt{1+u_x^2}-\varphi)^2-1} \right] \right\} & \quad u_x > u_c
	 \end{cases}
\end{equation}
with \(\gamma = \sqrt{1+ u_x^2 + u_y^2 + u_z^2 }\), \(\gamma_\perp = \sqrt{1+ u_y^2 + u_z^2 }\), \(\gamma_c = 1+ \varphi\) and \(u_c = \sqrt{\gamma_c^2-1}, \beta_c =  \sqrt{2\varphi}\) for \(\varphi \ll 1\) and
\begin{eqnarray}
	n_{r0} &=& \frac{\gamma_0^2\mu^2}{2 \pi e^\mu} \left[ 2u_c (1+\gamma_0 \mu) e^{-\mu \gamma_0} + e^{\mu \gamma_0 \varphi} \sum_{\pm} \int_{g_c}^\infty
	d \gamma \frac{\gamma (1+\mu \gamma_0 \gamma)}{\sqrt{\gamma^2-1}}  e^{[-\mu(\gamma_0 \gamma \pm u_0 \sqrt{(\gamma-\varphi)^2-1})]}\right]^{-1}
\end{eqnarray}
is used to calculate the dispersion relation
\begin{equation}
	k^2 c^2 - \omega^2 - \omega_{pe}^2 (U_e + V_e) = 0
\end{equation}
with the definitions
\begin{equation}
	U_e = \int_{-\infty}^\infty d^3u \frac{u_x}{\gamma} \frac{\partial f_{re}}{\partial u_x}
\end{equation}
and
\begin{equation}
	V_e = \int_{-\infty}^\infty d^3u \frac{u_x^2}{\gamma \left( \gamma \frac{\omega}{kc}- u_z \right)} \frac{\partial f}{\partial u_z}
\end{equation}
where only fluctuations \(\mathbf k = k \mathbf{e_z}\) perpendicular to the fluid velocity \(\mathbf{u_0} = u_0 \mathbf{e_x}\) were considered.
The parameters were derived to
\begin{eqnarray}
	U_r &=& - n_{r0} 2 \pi \mu e^{\mu (\gamma_0 \varphi +1 )} \sum_\pm \int_{\gamma_c}^\infty \sqrt{\gamma^2 -1} e^{\mp \mu u_0 \sqrt{(\gamma - \varphi)^2-1}} \nonumber \\
	&& \times \left[ \gamma_0 \gamma \Gamma \left( 0,\mu \gamma_0  \gamma \right) \pm \frac{\beta_0}{\mu} \frac{\gamma - \varphi}{\sqrt{(\gamma - \varphi)^2-1}} e^{-\mu \gamma_0 \gamma} \right]
\end{eqnarray}
and
\begin{eqnarray}
	V_r &=& n_{r0} 4 \mu \gamma_0 \int_0^\infty du_z u_z^2 \int_0^\infty du_y  \left[  \frac{2}{\gamma_\perp} e^{-\mu (\gamma_0 \gamma_\perp -1)}   \int_0^{u_c} d u_x \frac{u_x^2}{\gamma (\gamma^2 y^2 + u_z^2)}  \right. \nonumber \\
	&& + \left.  \sum_\pm e^{\mu (\gamma_0 \varphi + 1)} \int_{u_c}^\infty \frac{u_x^2}{\gamma^2 (\gamma^2 y^2 + u_z^2)} e^{-\mu( \gamma_0 \gamma \pm u_0 \sqrt{(\sqrt{1+u_x^2}-\varphi)^2-1}) } \right]
\end{eqnarray}
and solved numerically for the general case.

\end{document}